\documentclass[a4paper,12pt]{article}
\usepackage[cp1251]{inputenc}
\usepackage[english, russian]{babel}
\usepackage[dvips]{graphicx}
\graphicspath{{.}}
\usepackage{amsfonts}
\usepackage{amsmath}

\begin{document}
\mathsurround=2pt \sloppy
\title{Splitting of the superfluid transition in $^3$He by nematic aerogels.}
\author { I. A. Fomin
\vspace{.5cm}\\
{\it  P. L. Kapitza Institute for Physical Problems}\\ {\it Russian
Academy of Science},
\\{\it Kosygina 2,
 119334 Moscow, Russia}}

\maketitle
\begin{abstract}
Role of inhomogeneous perturbations of nematic aerogel on the form of the order parameter emerging at the transition of liquid $^3$He in the superfluid state is considered. It is shown that a region of stability of the polar distorted ABM phase can begin right from the transition temperature. The symmetry argument is given, which selects the most favorable aerogel for stabilization of pure polar phase.
\end{abstract}

\section{Introduction}
In the recent years new  superfluid phases of liquid $^3$He  were successfully stabilized \cite{Dm-pp,Dm-rev}. Existence of multiple phases is a typical manifestation of unconventional Cooper pairing. In case of $^3$He it is pairing in the state with the orbital moment $l=1$ and spin $s=1$. The corresponding order parameter is proportional to complex  3$\times$3 matrix $A_{\mu j}$, its first index enumerates 3 projections of spin, while the second index - three projections of the orbital moment. According to Landau theory transition occurs   when the coefficient at the second order invariant in the expansion of the gain of free energy in powers of the order parameter changes sign. In $^3$He the second order term is $\delta f\sim a(P,T)A_{\mu j}A_{\mu j}^\ast$.   The temperature of transition $T_c$ at a given pressure $P$
is determined by the condition $a(P,T_c)=0$. This temperature is ``degenerate'' in a sense that at this temperature the normal phase develops instability to formation of Cooper pairs  with all three possible projections of spin $s_z=0$, $s_z=\pm 1$ and three projections of orbital moment   $l_z=0$, $l_z=\pm 1$. The proper combination of these basis functions is found by minimization of the contribution of the fourth order invariants to the gain of the free energy. In $^3$He there are five such invariants: $I_1=A_{\mu j}A_{\mu j}A_{\nu l}^*A_{\nu l}^*$, $I_2=A_{\mu j}A_{\mu j}^*A_{\nu l}A_{\nu l}^*$, $I_3=A_{\mu j}A_{\nu j}A_{\mu l}^*A_{\nu l}^*$, $I_4=A_{\mu j}A_{\nu j}^*A_{\nu l}A_{\mu l}^*$, $I_5=A_{\mu j}A_{\nu j}^*A_{\mu l}A_{\nu l}^*$, they enter free energy as a combination $\sum_{s=1}^5 \beta_sI_s$ where  $\beta_1,..,\beta_5$ are phenomenological coefficients. Generally this combination has many extrema and several minima \cite{march,VW}, but at realistic values of coefficients  $\beta_1,...,\beta_5$ only two of the minima are realized. They correspond to  Anderson-Brinkman-Morel (ABM) and Balian and Werthamer (BW) phases \cite{VW}.

New phases are of different origin. They are stabilized by external forces, which lower spherical symmetry of normal phase and split the transition.
Symmetry with respect to rotations of spin is broken by magnetic field. The resulting new phases A$_1$, A$_2$  were studied before \cite{koj}, we                                      will not discuss it here. No bulk field is available for splitting of the transition in liquid  $^3$He over projections of orbital moment. Mechanical rotation would require too high  angular velocities. Aoyama and Ikeda \cite{AI} suggested to use for lowering of orbital symmetry oriented anisotropic impurities. They considered ensemble of impurities with the average cross-section for scattering of the Fermi quasiparticles $\sigma(\hat{k})\sim [1+\delta(\hat{k}\cdot \hat{z})^2]$, where $\hat{z}$ is the assumed direction of  anisotropy and $\hat{k}$ - the momentum transfer at the scattering.  Following argument of the theory of superconducting alloys \cite{AG1} they concluded that the transition temperature will  split  in two - $T_{c0}$ corresponding to projection $l_z=0$, and $T_{c1}$ - to projections $l_z=\pm 1$. If anisotropy constant $\delta$ is negative $T_{c0}>T_{c1}$ and for each pressure even for small $\delta$ there exist a finite interval of temperatures starting from the temperature of transition in the superfluid state where Cooper pairs have  $l=1$ and  $l_z=0$. This state corresponds to the polar phase of $^3$He. This proposal stimulated experiments aimed at stabilization of the polar phase (for a review cf.\cite{Dm-rev}).  As a globally anisotropic ensemble of impurities were used different types of the so-called \emph{nematic} aerogels. These aerogels are formed by straight and nearly parallel strands.  In the earliest experiments it was ``obninsk'' aerogel \cite{Askh1}. Its strands of diameters $d\approx 9 nm$ consist of amorphous AlOOH.  Anisotropy of this aerogel, as judged from the ratio of the mean free pathes of the Fermi quasi-particles along the strands and perpendicular to this direction  $l_{\|}/l_{\bot}\approx 1.4$  is much greater than that assumed in the calculations of Aoyama and Ikeda. So it came as a surprise that instead of the polar phase the polar distorted ABM-phase which is a superposition of all projections of the angular moment  was observed in these experiments. When in the further experiments \cite{Dm-pp} the anisotropy in  $^3$He was induced by another nematic aerogel with even greater anisotropy - \emph{nafen}, the polar phase was securely identified and confirmed \cite{Parp}. The goal of this paper is to use  the Landau theory of the second order phase transitions for finding out which characteristics of aerogel are favoring the particular form of the order parameter, emerging at the   transition.

\section{Local anisotropy}
In theoretical discussions of the phase diagram of superfluid $^3$He in aerogel \cite{Sauls,SF1,fom3} in a framework of the Landau theory aerogel was considered as a continuous media, anisotropic but not chiral. The term of the second order in the expansion of the gain of free energy in that case can be written as $\delta f\sim \Lambda_{jl}(P,T)A_{\mu j}A_{\mu l}^\ast$, where $\Lambda_{jl}(P,T)$ is \emph{real symmetric} matrix. This matrix can be diagonalized and has real diagonal elements $\tau_x(P,T), \tau_y(P,T), \tau_z(P,T)$. Directions of  its eigenvectors, which are real and mutually orthogonal  can be taken as orts of the coordinate system.  Formally there may be 3 transition temperatures, when one of the diagonal elements  changes sign. Real transition takes place at the highest of the three temperatures. In the sited above papers aerogel was assumed on the average axially symmetric, so that e.g.  $\tau_x(P,T)=\tau_y(P,T)$ and $z$ is the symmetry axis. Such  ``field'' description of aerogel leaves aside all effects originating from the discrete structure of aerogel, in particular a possible variation of the order parameter in a space between the strands.

According to Ref.\cite{Rainer} the strands of aerogel perturb the condensate in their neighborhood on a distance of the order of the coherence length $\xi_0$ which in the superfluid  $^3$He depending on a pressure varies in the interval 20$\div$80 nm. Reaction of the condensate on this perturbation extends for the temperature dependent coherence length $\xi(T)$ which in a vicinity of $T_c$ meets the condition  $\xi(T)\gg\xi_0$. In this vicinity perturbation, induced by aerogel can be treated as local and described by the following term in the Landau expansion of free energy in powers of the order parameter $f_{ag}=N(0)\Lambda_{jl}(\mathbf{r})A_{\mu j}A_{\mu l}^\ast$.  Here $N(0)$ is the density of states. Matrix $\Lambda_{jl}(\mathbf{r})$ depends on coordinate $\mathbf{r}$ and it is \emph{Hermitian} since the free energy has to be real. It is assumed that the strands of aerogel do not interact directly with spins of quasiparticles. In the experiments  \cite{Dm-rev} this interaction was eliminated by a thin film of  $^4$He covering the strands.
It is convenient to represent the matrix $\Lambda_{jl}(\mathbf{r})$ as the sum of its ensemble average: $\tau_{jl}=\langle\Lambda_{jl}\rangle$, and fluctuating local anisotropy  $\eta_{jl}(\mathbf{r})=\Lambda_{jl}(\mathbf{r})-\tau_{jl}$ so that  $\langle\eta_{jl}\rangle=0$.  In what follows we assume that the considered aerogels on the average are non-chiral, i.e. they do not distinguish left and right orientation. Then matrix $\tau_{jl}$ is real symmetric, its principal directions as was explained above determine orientation of coordinate axes $(x,y,z)$ and the principal values $\tau_x,\tau_y,\tau_z$ are real functions of temperature:$\tau_{jl}=\tau_x\hat{x}_j\hat{x}_l+\tau_y\hat{y}_j\hat{y}_l+\tau_z\hat{z}_j\hat{z}_l$
With these notations the expansion of the free energy density in a vicinity of temperature of instability of the normal phase can be written as:
$$
\frac{f_S-f_N}{N(0)}=[\tau_{jl}+\eta_{jl}(\mathbf{r})]A_{\mu j}^\ast A_{\mu l}+\xi_s^2\left(\frac{\partial A^*_{\mu
l}}{\partial x_n} \frac{\partial A_{\mu l}}{\partial
x_n}\right)+\frac{1}{2}\sum_{s=1}^5 \beta_sI_s.                                        \eqno(1)
$$
The contribution of the gradient terms is written here in a model isotropic form to avoid non-essential complications.  The equilibrium order parameter is found from the equation
$$
[\tau_{jl}+\eta_{jl}(\mathbf{r})]A_{\mu l}-\xi_s^2\left(\frac{\partial^2 A_{\mu
j}}{\partial x_n^2}\right)+\frac{1}{2}\sum_{s=1}^5 \beta_s\frac{\partial I_s}{\partial A_{\mu j}^\ast}=0.              \eqno(2)
$$
This system of differential equations hardly can be solved for a general form of  matrix $\eta_{jl}(\mathbf{r})$.  We limit ourselves to the case when the effect of random anisotropy $\eta_{jl}(\mathbf{r})$ can be treated as a perturbation. The perturbed solution of Eqn.(2) can be sought in a form $A_{\mu j}(\mathbf{r})=\bar{A}_{\mu j}+a_{\mu j}(\mathbf{r})$, where $\bar{A}_{\mu j}$ is the ensemble averaged order parameter.
This average, can be considered as the order parameter of a given phase of  $^3$He in aerogel, it describes thermodynamic properties of the liquid.
The fluctuating part  $a_{\mu j}(\mathbf{r})$ vanishes at such averaging.The condition of instability of the normal phase of liquid  $^3$He also has to be expressed directly in terms of  $\bar{A}_{\mu j}$. For this we substitute $A_{\mu j}=\bar{A}_{\mu j}+a_{\mu j}(\mathbf{r})$ in Eq. (2) and average the obtained expression.  Keeping terms up to the  second order over $a_{\mu j}$ and $\eta_{jl}(\mathbf{r})$ and linear with respect to  $\bar{A}_{\mu j}$ we arrive at:

$$
\tau_{jl}\bar{A}_{\mu l}+
\langle \eta_{jl}(\mathbf{r})a_{\mu l}(\mathbf{r})\rangle=0,                                           \eqno(3)
$$

The Greek index enters this equation as a parameter. In the experiments \cite{Dm-rev} all discussed phases belong to the ESP type, i.e. their order parameter can be factorized into   the spin and orbital parts. The spin part is a real vector $d_{\mu}$. Orbital anisotropy effect orientation of  $d_{\mu}$ only via weak dipole interaction, which is neglected here. In what follows we consider only orbital part of $A_{\mu j}$ and suppress the index $\mu$:
$$
\tau_{jl}\bar{A}_l+\langle \eta_{jl} a_l\rangle=0.                                                 \eqno(4)
$$
Fluctuating part $a_l$ is found from the linear equation
$$
\tau_{jl}a_l-\xi_s^2\left(\frac{\partial^2 a_j}{\partial x_n^2}\right)=-\eta_{jl}\bar{A}_ l.                   \eqno(5)
$$
It is solved by Fourier transformation
$$
a_l(\vec{k})=-(\tau_{ln}+\delta_{ln}\xi^2k^2)^{-1}\eta_{nm}(\vec{k})\bar{A}_m                                  \eqno(6)
$$
Substitution of this solution in
$$
\langle \eta_{jl} a_l\rangle=\int \eta_{jl}(-\vec{k})a_l(\vec{k})\frac{d^3k}{(2\pi)^3}                   \eqno(7)
$$
renders the linear equation for a temperature of instability of the normal phase:
$$
\left(\tau_{jl}-V_{jl}\right)\bar{A}_l=0,                                                                       \eqno(8)
$$
where
$$
V_{jl}=\left\langle\int\frac{d^3k}{(2\pi)^3}\eta_{mj}^{\ast}(\vec{k})(\tau_{mn}+\delta_{mn}\xi^2k^2)^{-1}\eta_{nl}(\vec{k})\right\rangle   \eqno(9)
$$
The relation $\eta_{jl}(-\vec{k})=\eta_{lj}^{\ast}(\vec{k})$ was used here. Eq. (8) has solution if
$$
det\left(\tau_{jl}-V_{jl}\right)=0.                                                              \eqno(10)
$$
Formally this is an eigenvalue problem with respect to the transition temperature $T=T_c$, which enters $\tau_{jl}$. For its solution one can apply the standard perturbation theory (e.g. \cite{LL5}). In zero order approximation $\tau_{jl}\bar{A}_l=0$ there are 3 solutions $\tau_x=0, \tau_y=0, \tau_z=0$ with corresponding eigenvectors $(1,0,0)$,  $(0,1,0)$ and
$(0,0,1)$. Assume that  $\tau_x \geq \tau_y > \tau_z$ so that the first transition occurs at $\tau_z(T=T_c)=0$. In a spirit of the Landau theory $\tau_z(T)$ can be expanded in the vicinity of $T_c$. With a proper normalization $\tau_z=(T-T_c)/T_c$ The first order correction to the transition temperature according to the recipe of perturbation theory is $(T-T_c)=T_cV_{zz}$. The emerging order parameter in this approximation has components  $\bar{A}_x=\frac{V_{13}}{\tau_x}$,  $\bar{A}_y=\frac{V_{23}}{\tau_y}$ and  $\bar{A}_z=1$. It follows from the definition (9)  that matrix $V_{jl}$ is Hermitian. It means that its symmetric part $(V_{jl}+V_{lj})/2$ is real and antisymmetric $(V_{jl}-V_{lj})/2$ - pure imaginary. If both $V_{13}$ and $V_{23}$ are real the transverse components can occur as a result of a small change of direction of $\bar{A}_l$. The emerging  phase is still polar, but with a different orientation. Imaginary increments to transverse components can not be included in the order parameter of polar phase. Together with the original $z$-component $\bar{A}_z=1$ they form the order parameter of the polar-distorted ABM-phase, which develops at further lowering of temperature in the ABM-phase continuously, without the phase transition. Pure polar phase will form if $V_{13}=V_{31}$ and $V_{23}=V_{32}$. This condition is met  for model forms of $\eta_{jl}(\vec{k})$ discussed in the literature \cite{AI,Sauls2}. These models are based on generalizations of the theory of superconducting alloys, representing aerogel as an ensemble of individual impurities.
In the discussed models $\eta_{jl}(\vec{r})$ is real symmetric tensor. That secures absence of imaginary part in  $V_{jl}$, but this restriction has no physical justification.
It is possible to formulate symmetry based restriction for idealized  nematic aerogels, i.e. for aerogels with infinitely long straight ideally parallel and smooth strands. Such aerogels have symmetry with respect to the mirror reflection in a plane normal to the strands.  In this case at the transformation $z\rightarrow(-z)$ components $V_{xz}$ and $V_{yz}$ of tensor $V_{jl}$ have to change their signs, but with the assumed symmetry it means that they have to be equal to themselves with the opposite sign, i.e. to vanish.
The aerogels nafen and mullite as is explained in Ref.\cite{Dm-rev} are more close to ideal than the ``obninsk''  samples. That can explain why the polar phase was observed in the first two and not observed in the last type of aerogel. It has to be mentioned that the ideal nematic aerogel does not suppress temperature of transition of  $^3$He in the superfluid state \cite{F18}. It means that the degree of suppression of the $T_c$ can be used as an indicator of quality of a given sample.

In conclusion, the average uniaxial anisotropy induced in the liquid  $^3$He by the immersed aerogel does not provide the sufficient condition for stabilization of the polar phase. Spatial fluctuations of the local anisotropy depending on a structure of aerogel can favor presence of imaginary components of the order parameter and formation at  $T_c$ of distorted ABM phase.  The most favorable for formation of pure polar phase is the ideal nematic aerogel.

\section{Acknowledgments}
I thank  V.V.Dmitriev, A.A. Soldatov , and A.N.Yudin for useful discussions and constructive criticism.

\end{document}